\documentclass[11pt,english]{article}
\usepackage{amssymb}
\usepackage{amsmath}
\usepackage{babel}
\usepackage{latexsym}
\usepackage{graphics}
\usepackage{amsfonts}
\usepackage{hyperref}

\allowdisplaybreaks[3]

\makeatletter
\@addtoreset{equation}{section}
\makeatother

\def\dalemb#1#2{{\vbox{\hrule height .#2pt
        \hbox{\vrule width.#2pt height#1pt \kern#1pt
                \vrule width.#2pt}
        \hrule height.#2pt}}}

\def\0{{\sst{(0)}}}
\def\1{{\sst{(1)}}}
\def\2{{\sst{(2)}}}
\def\3{{\sst{(3)}}}
\def\4{{\sst{(4)}}}
\def\5{{\sst{(5)}}}
\def\6{{\sst{(6)}}}
\def\7{{\sst{(7)}}}
\def\8{{\sst{(8)}}}
\def\n{{\sst{(n)}}}

\def\ep{\epsilon}
\def\td{\tilde}

\def\half{{\textstyle{\frac{1}{2}}}}

\def\qu{{\textstyle{1\over 4}}}

\let\a=\alpha  \let\g=\gamma \let\d=\delta 
    
 \let\m=\mu \let\n=\nu  \let\r=\rho
 \let\t=\tau    
\let\w=\omega

\def\nn{\nonumber} \def\bd{\begin{document}} \def\ed{\end{document}}
\def\ds{\documentstyle} \let\fr=\frac \let\bl=\bigl \let\br=\bigr
\let\Br=\Bigr \let\Bl=\Bigl
\let\bm=\bibitem
\let\na=\nabla
\let\pa=\partial \let\ov=\overline
\newcommand{\be}{\begin{equation}}
\newcommand{\ee}{\end{equation}}
\def\ba{\begin{array}}
\def\ea{\end{array}}
\def\ft#1#2{{\textstyle{{\scriptstyle #1}\over {\scriptstyle #2}}}}
\def\fft#1#2{{#1 \over #2}}
\def\del{\partial}
\def\sst#1{{\scriptscriptstyle #1}}
 \def\oneone{\rlap 1\mkern4mu{\rm l}}
\def\ie{{\it i.e.\ }}
\def\via{{\it via}}
\def\semi{{\ltimes}}
\def\str{{\rm str}}
\def\Dm{{{D_{\sst{max}}}}}
\def\vac{ \left | 0 \right \rangle }
\def\kvac{ \left | k \right \rangle }

\def\sp{\; \; \;}

\def\bol{ \left | B (p^+) \right \rangle}
\def\bo1{ \left | B^0 (p^+) \right \rangle}

\def\bolt{ \left | B (p^+) \right \rangle_{\t}}

\def\boxl{ \left | B (x^-) \right \rangle}

\def\<{ \langle }
\def\>{ \rangle }

\def\vf{\varphi}

\def\ls{{(l,0)}}
\def\lv{{(l,\pm1)}}
\def\lt{{(l,\pm2)}}

\def\lse#1{{(l_{#1},0)}}
\def\lve#1{{(l_{#1},\pm1)}}
\def\lte#1{{(l_{#1},\pm2)}}

\def\lsg#1{{5(l_{#1},0)}}
\def\lvg#1{{5(l_{#1},\pm1)}}
\def\ltg#1{{5(l_{#1},\pm2)}}

\def\lsi#1{{5{(#1,0)}}}
\def\lvi#1{{5{(#1,\pm1)}}}
\def\lti#1{{5{(#1,\pm2)}}}

\def\lsr#1{{1{(#1,0)}}}
\def\lvr#1{{1{(#1,\pm1)}}}
\def\ltr#1{{1{(#1,\pm2)}}}

\def\cD{{\cal D}}
\def\cE{{\cal E}}
\def\cF{{\cal F}}
\def\cG{{\cal G}}
\def\cH{{\cal H}}
\def\cK{{\cal K}}
\def\cO{{\cal O}}
\def\cP{{\cal P}}
\def\cQ{{\cal Q}}
\def\cR{{\cal R}}
\def\cS{{\cal S}}
\def\cT{{\cal T}}
\def\cU{{\cal U}}
\def\cV{{\cal V}}
\def\cW{{\cal W}}

\newcommand{\nono}{\nonumber}
\newcommand{\dtilde}[1]{\tilde{\tilde{#1}}}
\newcommand{\hatb}[1]{\hat{\ov{#1}}}
\newcommand{\hatt}[1]{\hat{\tilde{#1}}}
\newcommand{\emnr}{{e_\m}^{\n\r}}
\newcommand{\sub}[1]{\phantom{}_{(#1)}\phantom{}}

\newcommand{\comment}[1]{}

\def\hna{\hat{\na}}

\newcommand{\hsp}{\hspace{0.5cm}}

\newcommand{\ho}[1]{$\, ^{#1}$}
\newcommand{\hoch}[1]{$\, ^{#1}$}
\newcommand{\bea}{\begin{eqnarray}}
\newcommand{\eea}{\end{eqnarray}}
\newcommand{\ra}{\rightarrow}
\newcommand{\lra}{\longrightarrow}
\newcommand{\Lra}{\Leftrightarrow}
\newcommand{\ap}{\alpha^\prime}
\newcommand{\bp}{\tilde \beta^\prime}
\newcommand{\tr}{{\rm tr} }
\newcommand{\Tr}{{\rm Tr} }
\newcommand{\NP}{Nucl. Phys. }

\newcommand{\ams}{{\it Institute for Theoretical Physics,
University of Amsterdam, \\
Valckenierstraat 65, 1018XE Amsterdam, The Netherlands} \\
{\tt M.Taylor@uva.nl}}
\newcommand{\auth}{{\large Marika Taylor }}

\begin{document}
\begin{flushright}
\hfill{ITFA-2008-48}
\end{flushright}

\vspace{15pt}

\begin{center}

{\Large \bf Non-relativistic holography}

\vspace{20pt}

\auth

\vspace{15pt}

\vspace{8pt}

{\ams}

\vspace{15pt}

\underline{ABSTRACT}
\end{center}

We consider holography for $d$-dimensional 
scale invariant but non-Lorentz invariant field theories, which do 
not admit the full Schr\"{o}dinger symmetry group. We find new realizations of the
corresponding $(d+1)$-dimensional gravity duals, engineered with a variety of matter Lagrangians,
and their finite temperature generalizations. The thermodynamic properties of the finite temperature
backgrounds are precisely those expected for anisotropic, scale invariant field theories. The brane
and string theory realizations of such backgrounds are briefly discussed, along with their holographic
interpretation in terms of marginal but non Lorentz invariant deformations of conformal field theories. 
We initiate discussion of holographic renormalization in these backgrounds, and note that such systematic 
renormalization is necessary to obtain the correct behavior of correlation functions.

\pagebreak

\section{Introduction}

There has been considerable interest recently in studying holographic models for condensed 
matter physics. Motivated by fermions at unitarity, Son suggested exploration of holographic
duals to Galilean conformal field theories \cite{Son:2008ye}, and a number of subsequent papers have discussed
string theory realizations. Such theories can be defined in flat $(D+1)$-dimensional spacetime, with
coordinates $(t,x^i)$ where $i = 1, \cdots, D$, and exhibit anisotropic scale invariant behavior, namely
invariance under the scale transformation ${\cal D}$:
\be
x \rightarrow \lambda x; \qquad
t \rightarrow \lambda^{\eta} t,
\ee
where $\eta$ is the dynamical exponent. More generally, one could consider scale invariant
field theories in which the spatial coordinates also scale anisotropically, and holographic backgrounds
with such symmetry will indeed be mentioned
briefly here. Actually, for condensed matter applications, one typically is interested in 
considering anisotropy between different
spatial dimensions. 

In addition to the dilatation symmetry, the generic symmetry group will consist of 
time translations ${\cal H}$, spatial translations ${\cal P}^i$ and spatial rotations ${\cal M}^{ij}$. 
This symmetry algebra closes on itself; the algebra also closes on the 
addition of a single special conformal transformation ${\cal C}$. Addition of Galilean boosts ${\cal K}^i$ 
requires the further addition of a mass operator ${\cal M}$ to close the algebra; the resulting symmetry group
involving  time translations,
spatial translations, spatial rotations, Galilean boosts, dilatation, 
special conformal transformation and mass operator is the Schr\"{o}dinger symmetry group, which can be viewed as the 
non-relativistic version of conformal symmetry.

Much interest has been focused
on holographic duals of theories which admit the full Schr\"{o}dinger symmetry group 
\cite{Balasubramanian:2008dm,Goldberger:2008vg,Barbon:2008bg,Sakaguchi:2008ku,Wen:2008hi,Herzog:2008wg,Adams:2008wt,Maldacena:2008wh,Nakayama:2008qm,Chen:2008ad,Minic:2008xa,Pal:2008rf,Galajinsky:2008ig,Kovtun:2008qy,Duval:2008jg,
Lee:2008xf,Yamada:2008if,Sakaguchi:2008rx,Nakayama:2008qz,Lin:2008pi,Hartnoll:2008rs,Schvellinger:2008bf,Mazzucato:2008tr,
Rangamani:2008gi,Akhavan:2008ep}. 
To realize a theory with Schr\"{o}dinger symmetry in
$D+1$ dimensions holographically, it has been argued \cite{Son:2008ye} that 
one needs a dual $(D+3)$ dimensional geometry. As usual, one extra 
dimension is needed in the geometric dual to realize renormalization group flow, but in this case there
is a further extra dimension needed to realize the Schr\"{o}dinger symmetry group, in particular,
the Galilean boosts, special conformal transformation and mass operator. 

Suppose one is interested instead in theories which do not admit such Galilean boosts or a mass operator,
and therefore particle number is not conserved. Such theories have a number of condensed matter applications,
including optimally doped cuprates and non-fermi liquid metals near heavy electron critical points,
see for example the discussions in \cite{Con}. Motivated by observations of anisotropic scaling behavior, 
theoretical descriptions of anisotropic quantum critical points have been explored, both in the free field limit
and by lattice techniques. A particular focus of these explorations has been on the observed spatial locality of 
correlators at finite temperature and fixed time; such ultra-locality may underlie certain experimental 
observations. 

Given the diverse condensed matter applications, 
it would clearly be desirable to understand strongly coupled anisotropic scale invariant theories better. 
Holography can potentially provide a tool in this area, albeit with the usual caveat, that one can only access
large $N$ theories that admit weakly curved holographic duals. From a theoretical perspective, realizing 
holographic duals for such anisotropic but scale invariant theories would extend the set of gravitational backgrounds
which can be treated holographically, and this is an important aim in itself. Even before moving to applications
of such dualities, one would like to understand how they can be embedding into string theory, and to acquire 
structural evidence beyond the matching of specific quantities. The aim of this paper will be to take a few modest
steps in this direction. 

In the case without conserved particle number 
it is possible to find a $(D+2)$ dimensional background which realizes the appropriate 
symmetries, and to engineer such a background as a solution of a Lagrangian with reasonable matter fields, and
this was done in \cite{Kachru:2008yh}. The matter Lagrangian used in \cite{Kachru:2008yh} and in subsequent
generalizations \cite{SekharPal:2008uy,Pal:2008id} used vector and p-form fields coupled by Chern-Simons terms. 
Here we will demonstrate
other choices of matter Lagrangian which can be used to engineer the scale invariant anisotropic backgrounds.
The first, discussed in section \ref{s-one}, 
involves massive vector fields, and can be used to engineer both spatial and temporal anisotropy, whilst
the second, discussed in section \ref{s-two}, uses a massless scalar coupled appropriately to a gauge field. 

Both actions seem rather natural from a string theory perspective, although in this paper we will only briefly 
discuss brane and string theory realizations. They also have complimentary advantages. We present 
a finite temperature generalization of the anisotropic background in section \ref{s-two} as a solution of 
the second Lagrangian; the resulting black hole solution has precisely the correct thermodynamic properties to
correspond to an anisotropic scale invariant theory at finite temperature and will allow for holographic explorations
of transport properties etc. By contrast, we have not found a 
finite temperature generalization using the matter Lagrangian of section \ref{s-one} (or indeed that of  
\cite{Kachru:2008yh}); in section \ref{s-string} we will comment on the interpretation of the difficulty in finding 
such a solution.

On the other hand, the massive vector realization presented in section \ref{s-one} has the advantage that 
there exists a limit in which the background is an infinitesimal deformation of Anti-de Sitter. Therefore, one
can interpret the anisotropic background in terms of deformations of a relativistic conformal field theory; the
required deformation turns out to be marginal but non-Lorentz invariant, and we discuss this interpretation in
section \ref{s-def}. 

The matter Lagrangians used to engineer anisotropic scale invariant backgrounds are rather natural and generic,
and it is interesting to ask whether they can be obtained from consistent truncations of string theory 
compactifications and whether there exists any natural brane interpretation. A complete answer to these questions is
not yet known, but in section \ref{s-string} we will give various comments and suggestions about string theory
realizations. 

In the absence of a derivation of a holographic duality from a brane configuration, one can obtain strong
evidence that the duality exists if the bulk calculations reproduce the same analytic structure as those
in the non-relativistic conformal field theory. Matching the (asymptotic) symmetries will automatically ensure
that certain bulk quantities match those of the boundary theory, but the matching of the structure of
the bulk volume divergences with the UV divergences of the field theory is a highly non-trivial test of the duality.
It goes far beyond matching specific quantities. In this paper we will initiate 
development of holographic renormalization in these backgrounds, and in section \ref{s-scalar} we discuss
how such systematic renormalization is necessary to obtain the correct behavior of correlation functions. 
A companion paper \cite{mmt2} will develop holographic renormalization in detail for non-relativistic 
backgrounds. 

One of the main results of this paper is the finite temperature generalization of the anisotropic scale invariant 
background. This background can be used to explore the usual properties of interest, namely transport coefficients
and phase structure. The latter is briefly mentioned in section \ref{s-two}, but careful treatment of the former
requires systematic holographic renormalization, and will thus be left to \cite{mmt2}. A number of recent 
works have explored aspects of holography for scale invariant anisotropic systems, see for example 
\cite{Adams:2008zk} for
a discussion of higher derivative corrections and \cite{Horava:2008jf} for new examples of anisotropic 
but scale invariant gauge theories. The results in this paper should be of use in further 
developing holography for such theories, particularly at finite temperature.

\section{Backgrounds with anisotropic scale invariance: I} \label{s-one}

Let us consider metrics of the form
\be
ds^2 = dr^2 + \sum_{i=1}^{d} e^{2 \a_i r} \eta_{ij} dx^i dx^j,
\ee
such that the metric is $AdS_{d+1}$ in domain wall coordinates when all $\a_i = 1$.
The Ricci tensor of the metric is given by:
\be
R_{rr} = - \sum_{i} \a_i^2; \qquad
R_{ij} = - \eta_{ij} e^{2 \a_i r} \a_i \sum_{k} \a_k.
\ee 
Thus the tangent frame Ricci tensor components are diagonal, constant and negative 
definite when all the $\a_i$ are positive. As already discussed in \cite{Kachru:2008yh}, such metrics have
no curvature singularities, since the curvature invariants are finite constants. However, the spacetime
is geodesically incomplete, and infalling particles can experience large tidal forces as 
$r \rightarrow - \infty$. 

For these reasons, one might be concerned about trying to set up holography for such a background. However, 
in this paper 
we will find that there exists a finite temperature generalization of this background which has a regular horizon,
and for which the thermodynamic properties match those expected for an anisotropic scale invariant field 
theory. Moreover, we will show in \cite{mmt2} 
that a precise holographic dictionary can be set up. This dictionary implies
that the volume divergences of such spacetimes have the same analytic structure as those in a non-relativistic 
quantum field theory; in particular we will derive Ward identities
for correlation functions which match those implied by non-relativistic scale invariance. Furthermore, there are
no ambiguities in the bulk calculation of correlation functions. All calculations therefore reinforce 
the viewpoint that such backgrounds admit a sensible holographic interpretation. 

Let us now turn to the question of how such a background can be engineered. In \cite{Kachru:2008yh}, 
a four-dimensional anisotropic scale invariant background was engineered using 
an action involving a two form and a three form field with a Chern-Simons coupling. 
Such an action was argued to be rather generic in string theory, although no explicit brane realization or embedding
into ten-dimensions was given. Here we will consider other possibilities for the matter needed to support such a
background. Let us start by considering an action of the form:
\be
I = \frac{1}{16 \pi G_{d+1}}
\int d^{d+1}x \sqrt{g} \left (R  + \Lambda - \qu \sum_a ( (F^a)^2 + m_a^2 (A^a)^2) \right ), \label{one}
\ee
where $F^a_{mn} \equiv (\pa_m A^a_n - \pa_n A^a_m)$ are (Abelian) massive vector fields. 
The field equations are then:
\bea
R_{mn} &=& - \frac{\Lambda}{(d-1)} g_{mn} + \qu \sum_a 
\left ( 2 (F^a)_{mp} (F^a)_{n} ^{\;p} + m_a^2 (A^a)_m (A^a)_n \right ) \nn \\
&& - \frac{1}{4(d-1)} \sum_a (F^a)^2 g_{mn}. \\
D_{m} (F^a)^{mn} &=& \half m_a^2 (A^a)^n. \nn
\eea
Next suppose that the vectors are:
\be
(A^a)_i = {\cal A}_i \d^{a}_i e^{\beta_i r}, 
\ee
with ${\cal A}_i$ constants.
These satisfy the vector field equations provided that:
\be
2 \beta_i (\beta_i - \alpha_i + \sum_{j \neq i} \alpha_j ) = m_a^2.
\ee
By further restricting $\beta_i = \alpha_i$, the vectors contribute constant terms to the tangent 
frame components of the Ricci tensor. Under this restriction,
\be \label{mass}
2 \alpha_i \sum_{j \neq i} \alpha_j = m_a^2.
\ee
The Einstein equations then become:
\bea
R_{rr} &=& - \frac{\Lambda}{(d-1)} + \frac{(d-2)}{2 (d-1)} 
\sum_{i} \eta_{ij} (\a^i {\cal A}^i) (\a^j {\cal A}^j); 
\\
R_{ij} &=&  \eta_{ij} e^{2 \a_i r} \left ( - \frac{\Lambda}{(d-1)} - 
\frac{1}{2 (d-1)} \sum_{k} \eta_{kl} (\a^k {\cal A}^k) (\a^l {\cal A}^l) \right ) \nn \\
&& + \d_{ij} (\frac{1}{2} (\alpha_i {\cal A}_i)^2  + \frac{1}{4} m_a^2 {\cal A}_i^2) e^{2 \a_i r}. \nn
\eea
Before considering explicit solutions of these equations, let us note that 
for $d=4$ with only one vector field, this realization follows from integrating 
out the three form field of \cite{Kachru:2008yh}. The action used in the latter was:
\be
I = \int d^4 x \sqrt{g} (R + \Lambda  - F_{mn} F^{mn} - H_{mnp} H^{mnp}) - 2 c \int F \wedge B,
\ee
with $H = dB$ and $F = dA$. The resulting equations of motion for the matter fields are:
\be
d (\ast F) = - c H; \qquad d (\ast H) = c F,
\ee
which in turn implies that $F$ is massive, i.e.
\be
d (\ast d (\ast F) + c^2 A) = 0. 
\ee
Integrating the three form out reduces the matter terms in the action to those 
of a massive vector field. Both Chern-Simons couplings and massive vector fields generically
arise in string theory compactifications. Here we will find several advantages of 
engineering the anisotropic backgrounds with massive vector fields, notably a natural holographic interpretation
as a deformation of a conformal field theory, and a possible relation to known consistent truncations of
spherical compactifications.

\subsection{One anisotropic direction}

One can engineer a four-dimensional background of the form:
\be
ds^2 = dr^2 + (- e^{2 \alpha r} dt^2 + e^{2 \beta r} dx \cdot dx),
\ee
by considering an action with only one massive vector, with $m^2 = 4 \alpha \beta$. The
three independent Einstein equations can then be consistently solved with $A_t = {\cal A} e^{\alpha r}$ and
\be
{\cal A}^2 = 2 \a^{-1} (\a - \beta); \qquad
\Lambda = (\alpha^2 + \alpha \beta + 4 \beta^2).
\ee
Thus the vector field is real provided that $\alpha \ge \beta$. 
Such solutions lift immediately to analogous backgrounds in $(d+1)$-dimensions, with
\be
\Lambda = \a^2 + (d-2) \alpha \beta + (d-1)^2 \beta^2.
\ee
and ${\cal A}$ unchanged.

\subsection{Two anisotropic directions}

One can also engineer a $(d+1)$ dimensional background of the form:
\be
ds^2 = dr^2 + (- e^{2 \alpha r} dt^2 + e^{2 \beta r} dx \cdot dx_{d-2} + e^{2 \gamma r} dy^2),
\ee
with two massive vector fields such that
\be
A^0_t = {\cal A} e^{\alpha r}; \qquad
A^2_y = {\cal B} e^{\gamma r},
\ee
with masses satisfying (\ref{mass}) and
\bea
{\cal A}^2 &=& 2 \a^{-1} (\a - \beta); \\
{\cal B}^2 &=& 2 \gamma^{-1} (\beta - \gamma); \nn \\
\Lambda &=& \a^2 + \gamma^2 + (d-2) \beta (\alpha + \gamma) + \beta^2 (d-2)(d-1). \nn
\eea
The vector fields are real provided that $\alpha \ge \beta \ge \gamma \ge 0$. 

\subsection{General solution}

More generally, one can engineer a $(d+1)$-dimensional background:
\be \label{ads-kasner}
ds^2 = dr^2 + \left (- e^{2 \alpha r} dt^2 + e^{2 \beta_{a} r} dx^a \cdot dx^a \right ),
\ee
with $a = 1,\cdots,(d-1)$ and $\alpha \ge \beta_a $, $\beta_a \ge \beta_{a+1} \ge 0$ 
using $(d-1)$ massive vector fields such that:
\be
A^0_t = {\cal A} e^{\alpha r}; \qquad
A^a_a = {\cal B}_a e^{\beta_a r}, \qquad a \ge 2,
\ee
with masses satisfying (\ref{mass}) and
\bea
{\cal A}^2 &=& 2 \a^{-1} (\a - \beta_1); \\
{\cal B}_a^2 &=& 2 \beta_a^{-1} (\beta_{a-1} - \beta_{a}); \qquad a \ge 2;  \nn \\
\Lambda &=& \a^2 + \sum_{a} \beta_a^2 + (d-2) \beta_1^2 + (d-2) (\alpha \beta_1 + 
\sum_{a \ge 1}\beta_a \beta_{a+1} ). \nn
\eea
Clearly if any of the $\beta_a$ coincide the corresponding vector field $A^{a+1}$ vanishes. 
Note that some of the $\beta_a$ can vanish. In this case, the corresponding vector fields are 
massless, and the vector field is pure gauge. 

As an aside, it is interesting to observe that 
such backgrounds are somewhat analogous to (analytic continuations of) Kasner solutions. The metrics
for the latter are:
\be \label{lor-kasner}
ds^2 = d r^2 + \left ( - r^{2 \alpha} dt^2 + \sum_{a} r^{2 \beta_a} dx^a \cdot dx^a \right ).
\ee
The Ricci tensor components are:
\bea
R_{rr} &=& \frac{1}{r^2} \left ( (\alpha^2 + \sum_{a} \beta_a^2) - (\alpha + \sum_{a} \beta_a) \right ); \\
R_{tt} &=& - r^{2 (\alpha -1)} \alpha (\alpha + \sum_{a} \beta_a - 1); \nn \\
R_{ab} &=& - \d_{ab} r^{2 (\beta_a -1)} \beta_a \alpha + \sum_{a} \beta_a - 1). \nn
\eea
This is a solution of the vacuum Einstein equations provided that:
\be
(\alpha^2 + \sum_{a} \beta_a^2) = (\alpha + \sum_{a} \beta_a) = 1, 
\ee
which are the same conditions as in the more familiar Kasner solutions. 
The scale invariant backgrounds (\ref{ads-kasner}) have constant negative Ricci curvature everywhere, 
but there is geodesic incompleteness as $r \rightarrow - \infty$. The analytically continued 
Kasner solutions have zero Ricci curvature everywhere, but also have a timelike singularity at $r = 0$. 
It would be interesting to explore whether there is billiard behavior in the Lorentzian Kasner solutions, and indeed
whether there is any analogous structure in the scale invariant backgrounds. 

\subsection{Generalizations}

One can immediately generalize the previous discussion to involve massive $p$-forms, instead of
vector fields. For example, consider the Lagrangian:
\be
I = \frac{1}{16 \pi G_{d+1}} \int d^{d+1} x \sqrt{g} \left ( R + \Lambda - \frac{1}{6} (H_3^2 + m^2 B_2^2) \right),
\ee
in which $H_{mnp} = (\pa_{m} B_{np} + \cdots)$. This Lagrangian allows the following metric as a solution:
\be
ds^2 = dr^2 + e^{2 \alpha r} (-dt^2 + dy^2) + e^{2r} dx \cdot dx_{d-2},
\ee
provided that
\bea
\Lambda &=& \left ( d^2 + 3 \alpha d  - 4 d + 4 \alpha^2 - 8 \alpha + 4 \right ); \\
m^2 &=& 6 \alpha (d-2), \nn 
\eea
and the two form is
\be
B_{tx} = b e^{2 \alpha r}; \qquad b^2 = \frac{(\alpha-1)}{2 \alpha}. 
\ee
Such a background exists anisotropic scaling, with the dynamical exponents in the $t$ and $y$ directions
being the same. Clearly many further generalizations involving higher $p$-forms, and combinations of such massive
$p$-forms, should be possible. 

\section{Backgrounds with anisotropic scale invariance: II} \label{s-two}

Let us now consider a second type of matter Lagrangian that can be used to engineer 
backgrounds with anisotropic scale invariance. One takes the matter to be a massless scalar field,
coupled to a gauge field:
\be
I = \frac{1}{16 \pi G_{d+1}} 
\int d^{d+1}x \sqrt{g} \left (R  + \Lambda - \frac{1}{2} (\partial \phi)^2 - \qu e^{\lambda \phi} F^2 \right ), 
\label{two}
\ee
The field equations are then
\bea
R_{mn} &=& - \frac{\Lambda}{(d-1)} g_{mn} + \half (\partial_m \phi) (\partial_n \phi) 
+ \qu  e^{\lambda \phi} \left ( 2 F_{mp} F_{n} ^{\;p} \right ) \nn \\
&& - \frac{1}{4(d-1)} e^{\lambda \phi} F^2 g_{mn}. \\
D_{m} ( e^{\lambda \phi} F^{mn}) &=& 0; \qquad \Box \phi = \qu \lambda e^{\lambda \phi} F^2. \nn
\eea
Then the metric:
\be
ds^2 = dr^2 - e^{2 \alpha r} dt^2 + e^{2r} dx \cdot dx_{d-1}, \label{down}
\ee
can be engineered with the following dilaton and field strength:
\bea
F_{rt} &=& f e^{(\alpha + d - 1)r}; \qquad e^{\lambda \phi} = \mu e^{2 (1-d) r}; \\
\lambda^2 &=& 2 \frac{d-1}{\alpha - 1}; \qquad \Lambda = (\alpha + d-1) (\alpha + d-2); \nn \\
\mu f^2 &=& 2 (\alpha-1) (\alpha + d -1). \nn
\eea
When $\alpha =1$, the dilaton should be constant and the field strength vanishing; the limit $\alpha \rightarrow 1$ is
not smooth as the coupling $\lambda$ of the dilaton to the gauge field diverges in this limit. Reality of
the dilaton coupling to the gauge field requires that $\alpha > 1$; therefore one engineers only theories
in which there is critical slowing down, rather than speeding up. 

Note that there is effectively a free parameter, $\mu$, in this solution, which arises from the shift symmetry of
$\phi$ in the Lagrangian. In the case of $\alpha =1$, when the dilaton is constant, the usual AdS/CFT dictionary
relates the constant value of the dilaton to the coupling constant in the dual theory. We will see in \cite{mmt2}
that the dilaton here plays an analogous role: it also sources a coupling constant in the dual theory. By contrast, the
the gauge field is fixed entirely by the metric and the scalar field, and does {\it not} source any operator in the dual
theory. 

For $\alpha \neq 1$, and fixing the parameters $(\Lambda, \lambda)$ in the action so that the equations of motion
admit the anisotropic solution, there is also an $AdS_{d+1}$ solution of the field equations such that
\bea
ds^2 &=& L^2 \left ( dr^2 + e^{2r} (-dt^2 + dx \cdot dx_{d-1}) \right ); \label{up} \\
L^2 &=& \frac{d (d-1)}{(d + \alpha -1) (d + \alpha -2)}. \nn 
\eea
with $\phi$ constant and $F = 0$. The corresponding finite temperature solution is
\be
ds^2 = L^2 \left ( \frac{dr^2}{(1 - \bar{m} e^{-3r})}  + 
e^{2r} (- (1 - \bar{m} e^{-dr})dt^2 + dx \cdot dx_{d-1}) \right ),
\ee
with the horizon being located at
\be
r_h = \frac{1}{d} \ln (\bar{m}).
\ee
It is also interesting to observe that there is a generalization to an anisotropic 
finite temperature solution for this Lagrangian. Consider the metric:
\be 
ds^2 =  \frac{dr^2}{(1 - m e^{- (\alpha + d -1) r})} - (1 - m e^{ - (\alpha + d -1) r}) 
e^{2 \alpha r} dt^2 + e^{2r} dx \cdot dx_{d-1}. \label{brane}
\ee
This metric also satisfies the field equations, with the same field strength and dilaton as in
the zero temperature solution. This background has a regular horizon at $r_{h}$ with
\be
r_h = \frac{1}{(\alpha + d -1)} \ln (m),
\ee
with the entropy being:
\be
S = m^{\frac{d-1}{\alpha + d -1}} \frac{V_{d-1}}{4 G_{d+1}},
\ee
where $V_{d-1}$ is the volume of the transverse directions. The Hawking temperature is:
\be
T_{H} = \frac{1}{4 \pi} (\alpha + d -1) m^{\frac{\alpha}{\alpha + d-1}},
\ee
and therefore the entropy can be rewritten as:
\be
S = \frac{V_{d-1}}{4 G_{d+1}} \left ( \frac{4 \pi}{(\alpha + d -1)} \right )^{\frac{d-1}{\alpha}} 
T_{H}^{\frac{(d-1)}{\alpha}}.
\ee
This expression is manifestly of the form expected for a relativistic conformal field theory 
when $\alpha = 1$. In the next section we will discuss how this matches with expectations for
an anisotropic scale invariant theory. 

Note that the finite temperature solution (\ref{brane}) is not a solution to the equations of motion
following from (\ref{one}). The metric (\ref{brane}) is static and preserves $(d-1)$-dimensional 
rotational and translational 
symmetry. The only ansatz for the massive vector field which is consistent with these symmetries
is $A_t = A_t(r)$; $A_r = A_r (r)$ and $A_{i} = 0$. The vector field equation immediately sets $A_r = 0$, whilst
the $(t,t)$ and $(i,i)$ components of the Einstein equations imply that:
\be
(A_t)^2 = \frac{2}{\alpha} (\alpha - 1) | g_{tt} |; \qquad
F^2 = - 4 \alpha (\alpha -1). 
\ee
These constraints are however incompatible with the remaining Einstein equation, and the vector field equation,
except when $m=0$. Thus a finite temperature (static, symmetric) 
solution for this Lagrangian would need to take a different form; we will comment later in section \ref{s-string}
on the possible reasons for this.  

One can also compute the other thermodynamic parameters for the finite temperature 
solution. The Euclidean action, including the Gibbons-Hawking boundary term, is:
\be \label{euc-act}
I_E = - \frac{1}{16 \pi G_{d+1}} 
\int d^{d+1}x \sqrt{g} \left (R  + \Lambda - \frac{1}{2} (\partial \phi)^2 - \qu e^{\lambda \phi} F^2 \right ) - 
\frac{1}{8 \pi G_{d+1}}  \int d^{d} x \sqrt{h} K. 
\ee
(Note that the electric field is imaginary when analytically continued to the Euclidean section.)
The onshell action for the finite temperature solution is then:
\be
I_{E} = - \frac{m V_{d-1} \beta_H}{16 \pi G_{d+1}}.
\ee
This can be evaluated either using background subtraction relative to the zero temperature solution, or by
anticipating the counterterm action which will be derived elsewhere \cite{mmt2}. 
In the case at hand the only contributing counterterm is \cite{mmt2}:
\be
I_{ct} = (\alpha + d -2) \frac{1}{8 \pi G_{d+1}} \int d^d x \sqrt{h}. 
\ee
Note that there is no finite contribution to the action as $m \rightarrow 0$. 
The solution also carries a conserved electric charge, which is given by
\be
Q_e = \frac{1}{32 \pi G_{d+1}} \int e^{\lambda \phi} (\ast F) = \frac{\mu f}{32 \pi G_{d+1}} V_d,
\ee
with the potential at the horizon being:
\be
\Phi = A_{t}(r_h) = \frac{fm}{(\alpha + d -1)}. 
\ee
The mass is 
\be
M = \frac{m V_{d-1} (d-1)}{16 \pi G_{d+1}}. 
\ee
and thus one finds that
\be
I_{E} = \beta (M - Q_e \Phi) - S,
\ee
as one expects for the thermodynamic relation. 
Again the mass can be obtained either by anticipating the expressions for the renormalized one point functions
\cite{mmt2} or by subtraction from the zero temperature background. In the latter method, 
one evaluates the Komar integral
\be
M = - \frac{1}{8 \pi G_{d+1}} \oint d S_{mn} D^{m} k^{n},
\ee
where $k = \pa_{t}$, subtracting the zero temperature background to remove the infinite volume divergence. One obtains
precisely the same answer using the renormalized one point functions; in this case, the one point functions for 
dual operators $({\cal T}, {\cal T}_{ij})$ 
can be expressed in terms of the coefficients in the asymptotic expansion of the metric  
as \cite{mmt2}:
\be \label{vev2}
\< {\cal T} \> = \frac{1}{16 \pi G_{d+1}} (d + \alpha -1 ) g_{(d + \eta) tt} + \cdots; \qquad
\< {\cal T}_{ij} \> = \frac{1}{16 \pi G_{d+1}} (d + \alpha - 1 )g_{(d + \eta) ij} + \cdots,
\ee
where the metric in Fefferman-Graham type coordinates is expanded as 
\be
ds^2 = \frac{d \rho^2}{4 \rho} + \frac{1}{\rho} \left ( \frac{dt^2} {\rho^{\eta/2}} 
(- 1 + \rho^{\frac{d + \eta}{2}}  g_{(d + \eta) tt}) + dx^{i} dx^j (\delta_{ij} + 
\rho^{\frac{d + \eta}{2}}  g_{(d + \eta)ij})  + \cdots \right ),
\ee
and $\alpha = 1 + \eta$. Note that in the case of $\eta = 0$ these expressions reduce to
the expressions given for asymptotically $AdS_{d+1}$ spacetimes in \cite{deHaro:2000xn}. The ellipses
in (\ref{vev2}) denote terms non-linear in the sources; for the asymptotically $AdS_{d+1}$ case
these are given in \cite{deHaro:2000xn}, and for the case of interest here they will be given in \cite{mmt2}.
They do not, however, contribute in the case at hand, where the sources are constant. 
Evaluating the expressions (\ref{vev2}) for the black hole solution, one finds that
\be
\< {\cal T} \> = \frac{m}{16 \pi G_{d+1}} (d-1); \qquad
\< {\cal T}_{ij} \> = \frac{m}{16 \pi G_{d+1}} \d_{ij}.
\ee
Evaluating the mass gives
\be
M = \int d^{d-1} x \<{\cal T} \> = \frac{m}{16 \pi G_{d+1}} (d-1) V_{d-1}.
\ee
Note that there is a dilatation Ward identity relating the one point functions:
\be
\< {\cal T} \> = \d^{ij} \< {\cal T}_{ij} \>, 
\ee
which as will be discussed in section \ref{s-field} is the expected Ward identity for
an anisotropic but scale invariant theory. 

\bigskip

It is interesting to explore possible phase transitions at finite temperature, as was
done in the AdS case in \cite{Witten:1998zw}. Here there is no interesting phase structure for
the anisotropic black brane relative to the thermal anisotropic background (i.e. the zero temperature
background with time compactified), as the latter has zero action. However,
suppose one compares the free energy of the anisotropic black brane background, with that of
the finite temperature AdS black brane, at the same temperature. The free energy ${\cal F}$ is given
by $I_{E} = \beta {\cal F}$ and is 
\be
{\cal F} = - \frac{1}{16 \pi G_{d+1}} V_{d-1} (4 \pi (\alpha + d -1))^{1 + \frac{d-1}{\alpha}} 
T_{H}^{1 + \frac{d-1}{\alpha}}.
\ee
Comparing the free energies between backgrounds with $\alpha \neq 1$ and the $\alpha = 1$ AdS background, 
one finds that the anisotropic background is favored at low temperature, with the AdS background favored at high
temperature. The critical temperature is:
\be
T_c = \frac{1}{4 \pi} \left ( \alpha + d -1 \right )^{\frac{\alpha + d -1}{(d-1)(\alpha -1)}} 
d^{- \frac{d \alpha}{(d-1)(\alpha -1)}},
\ee
which for $d=3$, $\alpha =2$ is:
\be
T_{c} = \frac{4}{27 \pi}.
\ee
It would be interesting to investigate whether there exist any experimental results which reflect
such phase structure.

\section{Anisotropic scale invariant deformations} \label{s-def}

Consider the case of a background with one anisotropic direction such that:
\be
ds^2 = dr^2 + (- e^{2 \alpha r} dt^2 + e^{2r} dx \cdot dx),
\ee
which is supported by a massive vector field of mass $m^2 = 2 \alpha (d - 1)$ such
that
\be
A_t = {\cal A} e^{\alpha r}; \qquad {\cal A}^2 = \frac{\alpha -1}{2 \alpha}. 
\ee
Now let us consider the case where $\alpha = 1 + \eta$ and expand in $\eta$:
\bea
ds^2 &=& dr^2 + (- e^{2 r} (1 + 2 \eta r + \cdots )dt^2 + e^{2r} dx \cdot dx); \\
A_t &=&  \sqrt{\eta/2} e^r (1 + {\cal O}(\eta) + {\cal O} (\eta r) ); \nn \\ 
m^2 &=& 2 (d-1) + {\cal O}(\eta). \nn
\eea
Then to leading order in $\eta$ the background is $(d+1)$-dimensional AdS space with
a massive vector field perturbation. Solving this vector field equation in the 
fixed AdS background one finds that:
\bea
A_r &=& e^{(\zeta_{-} -2) r}(A(x,t) + \cdots) + e^{(\zeta_{+} -2) r}(\td{A}(x,t) + \cdots); \\
A_{\mu} &=& e^{\zeta_{-} r}(A_{\mu}(x,t) + \cdots) + e^{\zeta_{+} r}(\td{A}_{\mu}(x,t) + \cdots), \nn
\eea
where the ellipses denote expansions in powers of $e^{-2 r}$, with
\be
\zeta_{-} = 1; \qquad \zeta_{+} = (1 - d),
\ee
and 
\be
A(x,t) = \frac{1}{(d-1)} \pa_{\mu} A^{\mu}(x,t); \qquad
\td{A}(x,t) =  - \pa_{\mu} \td{A}^{\mu}(x,t). 
\ee
Then $A_{\mu}(x,t)$ is the source for the dual vector operator ${\cal O}_{\mu}$, 
which has (marginal) scaling dimension $d$. At the linearized level, one can add marginal
deformations to the conformal field theory:
\be
{\cal L} \rightarrow {\cal L} + \int d^{d} x a^{\mu} {\cal O}_{\mu}.
\ee
Such a deformation is consistent with scale invariance, but breaks the $d$-dimensional Poincar\'{e} symmetry. 
Comparing with the linearized solution, one sees that the dual conformal field theory is in this case 
deformed by the marginal vector
operator ${\cal O}_{t}$, with the deformation parameter being proportional to $\sqrt{\eta}$:
\be
{\cal L} \rightarrow {\cal L} + \int d^{d} x \sqrt{\eta} {\cal O}_{t}. 
\ee
This deformation preserves the symmetry group of the spatial directions. 
Generically such a deformation will not preserve scale invariance to higher order in $\eta$; one would need to 
demonstrate this explicitly. However, the fact that
there exists dual backgrounds at finite $\eta$ which have scale invariance suggests that there indeed exist such 
anisotropic scale invariant quantum field theories, related to the original CFTs by such deformations. 

In the context of the ${\cal N} = 4$ SYM duality, one can immediately identify possible marginal
vector deformations using the table related Kaluza-Klein supergravity modes to CFT operators reviewed in
\cite{D'Hoker:2002aw}. Indeed, already from the spectrum computed in \cite{Kim:1985ez}, one sees that the
only vectors in $AdS_5$ with the correct mass are those arising from coupled metric/five-form 
fluctuations. Let $g_{mn} = g^{o}_{mn} + h_{mn}$ where $g^{o}_{mn}$ is the background $AdS_5 \times S^5$ metric,
and let the four form potential be $C_{mnpq} = C^{o}_{mnpq} + c_{mnpq}$, where $C^{o}_{mnpq}$ is the 
background value. Now consider vector fluctuations such that:
\be
h_{\mu a} = B^{k I_v}_{\mu} (x) Y^{k I_v}_a (y); \qquad C_{\mu a b c} = \phi^{k I_v}_{\mu} (x) 
\epsilon_{abc de} D^d (Y^{k I_v})^e (y),
\ee 
where $x$ and $\mu$ are $AdS_5$ coordinates and indices respectively, with $y$ and $a$ $S^5$ coordinates and 
indices. Here the $Y^{k I_v}_a$ are vector spherical harmonics of degree $k$, which satisfy 
\be
\Delta Y^{k I_v} = - (k+1) (k-3) Y^{k I_v} \qquad k \ge 1,
\ee
where $\Delta$ is the Hodge-de Rham operator and $I_v$ denotes the residual $SO(6)$ representation labels. 
From  \cite{Kim:1985ez} one sees that the combination 
\be
{\cal B}_{\mu} \equiv ( B^{k I_v}_{\mu} - 4 (k+3) \phi^{k I_v}_{\mu}) 
\ee
diagonalises the equations of motion and has mass in $AdS_{d+1}$ (in our conventions) of
\be
m^2_{\cal B} = 2 (k^2 - 1). 
\ee
Therefore the $k=2$ modes, from the ${\bf 64}$ representation of $SO(6)$, are dual to marginal vector operators. 
The corresponding SYM operators (see \cite{D'Hoker:2002aw}) are
\be
{\rm Tr} (\lambda \bar{\lambda} X),
\ee
where $X$ denotes the six ${\cal N} = 4$ SYM scalars and $\lambda$ denotes the Weyl fermions. Whilst there has
been considerable exploration of (Lorentz invariant) marginal deformations of ${\cal N} = 4$ SYM, such marginal
but anisotropic deformations have not been discussed. It would be interesting to explore such deformations 
perturbatively, and check whether the deformation remains marginal to next order in the perturbation parameter 
$\eta$. 

\bigskip

Now let us turn to the case where the matter sector consists of the massless scalar coupled to
a gauge field. As mentioned already in (\ref{up}), this Lagrangian admits an $AdS_{d+1}$ solution, with
the massless scalar and gauge field corresponding to a marginal scalar operator and global currents of dimension 
$(d-1)$ respectively, in the dual $d$-dimensional CFT. The coupling $\lambda$ in the Lagrangian (\ref{two}) will 
determine the three point function between these operators. The same Lagrangian also admits the anisotropic scale
invariant solution (\ref{down}), and 
one might wonder whether one can interpret the dual theory as a deformation of a CFT, along
the lines of the arguments given above. However, in this case, the limit $\alpha \rightarrow 1$ is not smooth, so
one cannot expand perturbatively about $\alpha = 1$. Without a small expansion parameter, one cannot interpret 
(\ref{down}) in terms of a linearized deformation of a conformally invariant background. The anisotropic solution 
(\ref{down}) can presumably be interpreted in terms of a non-linear deformation of a conformally invariant background,
and this interpretation might become clearer if one identifies a string realization of this Lagrangian. 

\section{Embedding into string theory} \label{s-string}

In this section we will discuss supergravity and brane realizations of these anisotropic scale-invariant backgrounds. 

\subsection{Supergravity realizations}

We have realized anisotropic scale invariant backgrounds as solutions of an Einstein action with a cosmological 
constant and a matter content consisting of massive $p$-form fields or a massless scalar coupled to a gauge field. 
One would like to find cases where actions of this type can be obtained from consistent truncations of reductions of
ten-dimensional supergravity equations; any such examples should have a natural string theory embedding, and possibly
a brane interpretation.

Let us begin with the action involving a massive vector field (\ref{one}). Such an action was already
used in \cite{Son:2008ye} to engineer backgrounds exhibiting Schr\"{o}dinger symmetry. These backgrounds were
later embedded into ten-dimensional supergravity in \cite{Maldacena:2008wh}, where consistent sphere reductions
retaining massive vector fields were presented. It might seem as though the five-dimensional 
backgrounds of interest here can also be embedded into ten-dimensional supergravity, using this reduction. This
is unfortunately not the case, the principle reason being that the vector field used here is timelike 
(with $F^2 \neq 0$) whilst in the Schr\"{o}dinger examples the vector field was null. To illustrate this, let
us consider one of the consistent truncations given in \cite{Maldacena:2008wh}:
\bea
I_{5} &=& \frac{1}{16 \pi G_5} \int d^5 x \sqrt{g} \left (R + 24e^{-u-4v} - 4e^{-6u-4v} - 8e^{-10v} 
- 5 \partial u \partial u - \partial v \partial v \right . \\
&& \qquad \left . 
- \frac{1}{2} \pa \Phi \pa \Phi - e^{- \Phi + 4u + v} F_{ab} F^{ab} - 4 e^{-\Phi - 2u - 3v} A_a A^a \right ) \nn
\eea
This Lagrangian admits an $AdS_5$ vacuum solution in which all three scalar fields and the vector field are zero;
with respect to this vacuum the field $\Phi$ is massless, with the fields $(u,v,A_a)$ all massive. 
Since the Lagrangian contains a mass term for the vector, and a potential, one might think that the scale invariant 
anisotropic solutions of interest here can solve the equations of motions, with only the vector and metric non-zero.
This however is inconsistent with the scalar equations of motion:
\bea
\Box \Phi &=& - \frac{1}{4} e^{- \Phi + 4u + v}  F_{ab} F^{ab} - 4 e^{-\Phi - 2u - 3v} A_a A^a; \\
10 \Box u &=& 24 (e^{-u-4v} - e^{-6u-4v} ) + e^{- \Phi + 4u + v}  F_{ab} F^{ab} 
- 8 e^{-\Phi - 2u - 3v} A_a A^a; \nn \\
15 \Box v &=& 16(6e^{-u-4v} - e^{-6u-4v} - 5e^{-10v} )
+ \frac{1}{4} e^{- \Phi + 4u + v}  F_{ab} F^{ab} - 12 e^{-\Phi - 2u - 3v} A_a A^a. \nn
\eea 
These equations do not admit solutions in which $\Phi = u = v = 0$, with $F_{ab} F^{ab}$ and $A_a A^a$ negative
constants. By contrast, the scalar equations could be trivially solved in the Schr\"{o}dinger examples at 
zero temperature as the vector field was null: $F_{ab} F^{ab} = A_a A^a = 0$. In the finite temperature 
Schr\"{o}dinger backgrounds, the scalar field profiles are however non-trivial \cite{Maldacena:2008wh}. 

Note also that the truncations given in \cite{Maldacena:2008wh} involve massive vectors dual to vector
operators of dimension $\Delta =2 + \sqrt{3}$ and $\Delta = 2 + \sqrt{5}$, i.e. relevant and irrelevant operators
rather than the marginal operator of interest here. It seems reasonable to postulate that there exist analogous
consistent truncations which retain the vector fields dual to marginal operators, perhaps 
along with appropriate scalars. Recall also that we did not find a finite temperature generalization using the
massive vector Lagrangian; perhaps, as in the Schr\"{o}dinger case, one also needs to excite a non-trivial 
scalar field profile in the finite temperature solution.  
       
\bigskip

Now let us turn our attention to the other Lagrangian (\ref{two}) involving a massless scalar coupled to
a gauge field:
\be
I = \frac{1}{16 pi G_{d+1}} \int d^{d+1} x \sqrt{g} \left ( R + \Lambda - \frac{1}{2} (\pa \phi)^2 
- \frac{1}{4} e^{\lambda \phi} F^2 \right ).
\ee
Again at first sight one might think that such a Lagrangian could easily be obtained from well-known consistent 
truncations of string compactifications. For example, when $\Lambda = 0$, and for specific choices of $\lambda$,
this Lagrangian is a truncation of a toroidal reduction of Einstein gravity. For $\phi = F = 0$, the Lagrangian
can be obtained from consistent truncations of sphere reductions. The complete Lagrangian is superficially similar
to a truncation of a gauged supergravity theory, but there is one key difference. In the latter, the constant
cosmological constant is replaced by a scalar potential:
\be
I_{g} = \frac{1}{16 pi G_{d+1}} \int d^{d+1} x \sqrt{g} \left ( R + V(\phi) - \frac{1}{2} (\pa \phi)^2 
- \frac{1}{4} e^{\lambda \phi} F^2 \right ),
\ee 
with $V'(\phi) \neq 0$. The scalar field equation is therefore modified to:
\be
\Box \phi = - V'(\phi) + \frac{\lambda}{4} e^{\lambda \phi} F^2.
\ee              
The background of interest here 
is not a solution of these field equations, as $V'(\phi)$ in a gauged supergravity theory
is non-vanishing for any linear dilaton profile. Note also that typically in gauged supergravity theory the scalars
coupling to the gauge field are massive in the $AdS$ vacuum, which is not the behavior in the Lagrangian given above. 
At the same time, we found finite temperature generalizations solving these equations of motion, which had the 
expected thermodynamic properties, and thus the idea that such Lagrangians can be obtained from consistent truncations is 
quite compelling.

\subsection{Supersymmetry}

One might also wonder whether the gravitational backgrounds can be supersymmetric, and thence stable. Even without
an explicit embedding, it seems likely that the backgrounds can be embedded into a supersymmetric theory. 
A simple argument is the following. The symmetry group of the $d$-dimensional field theory 
consists of the dilatation symmetry ${\cal D}$, 
time translations ${\cal H}$, spatial translations ${\cal P}^i$ and spatial rotations ${\cal M}^{ij}$, satisfied
standard commutation relations with the dilatation acting as:
\be
[ {\cal D}, {\cal H} ] = - i \alpha {\cal H}; \qquad
[ {\cal D}, {\cal P}^i ] = - i {\cal P}^i. 
\ee
One can then ask whether this symmetry algebra can be extended into a supersymmetry algebra; for definiteness 
let us focus
on the case of $d=3$. Using a Majorana representation for the gamma matrices such that $\g^{0} = - i \sigma_2$;
$\g^1 = \sigma_3$ and $\g^2 = \sigma_1$, Majorana spinors can be expressed as
\be
Q = \left ( \begin{array}{c} q \\ - i q^{\ast}  \end{array}  \right ). 
\ee 
Then one can extend the symmetry algebra into symmetry algebra by the addition of a complex supercharge $q$
satisfying:
\be
[ {\cal J},q] = \frac{i}{2} q; \qquad  [ {\cal D}, q ] = - \frac{i}{2} \alpha q; \qquad
\{ q, q^{\ast}\} = {\cal H}, 
\ee
where ${\cal J} \equiv {\cal M}^{12}$ and all other commutators involving
 the supercharge are trivial. In the case that 
the dynamical exponent $\alpha = 2$, this superalgebra is a subalgebra of a super-Schr\"{o}dinger algebra.
Indeed, more generally, there has been considerable work on Schr\"{o}dinger superalgebras, 
see \cite{Leblanc:1992wu,Duval:1993hs,Henkel:2005dj,Sakaguchi:2008ku,Sakaguchi:2008rx,Nakayama:2008qz}, 
and these superalgebras generically admit
subalgebras of the type of interest here, 
in which there is no superconformal supersymmetry generator or special conformal transformation. 
Therefore, we would anticipate that backgrounds with anisotropic scale invariance can be 
supersymmetrized, and can be found within the framework of supersymmetric classifications. Namely, one
could consider classifications of supersymmetric solutions 
such as that given for minimal five-dimensional supergravity in \cite{Gauntlett:2002nw}, make an appropriate
anisotropic ansatz, and look for possible solutions.  

\subsection{Brane realizations}

By now the holographic realization of Schr\"{o}dinger symmetry is rather well understood, in terms of Melvin or TsT 
transformations of known dualities, see for example 
\cite{Herzog:2008wg,Maldacena:2008wh,Kovtun:2008qy,Lin:2008pi,Hartnoll:2008rs,Schvellinger:2008bf,Mazzucato:2008tr}. 
One would anticipate that the geometries of interest here can also be engineered with branes and one avenue to explore
has already been mentioned: the anisotropic and marginal deformations of conformal field theories could correspond to 
placing D3-branes in a non-trivial background in which Lorentz invariance has been broken. Another avenue which seems interesting (but
as yet not yielded a concrete implementation) is the following. The string frame metric for a stack of Dp-branes is
given by:
\be
ds^2 = H^{-1/2} dx \cdot dx_{p+1} + H^{1/2} dy \cdot dy_{9-p},
\ee
where $H(y)$ is a harmonic function in the $(9-p)$ transverse directions $y$. Supersymmetric brane intersections are constructed
according to the standard rules; now suppose that one has an intersection in which the metric was 
\bea
ds^2 &=& - (h_1 h_2 h_3 h_4)^{-1/2} dt^2 + (h_1 h_2 h_3)^{-1/2} (h_4)^{1/2} dx \cdot dx_2 + (h_1 h_2 h_3 h_4)^{1/2} ds^2(R^3) 
\nn \\
&& + (h_1 h_2)^{-1/2} (h_3 h_4)^{1/2} dy \cdot dy_2 + + (h_1 h_3)^{-1/2} (h_2 h_4)^{1/2} dz \cdot dz_2, \label{int}
\eea
with all four functions $h_a$ being harmonic and single centered in $R^3$, i.e. $h_a = (1 + q_a/r)$. Taking the decoupling 
limit via $r \rightarrow 0$ by construction gives rise to a background with anisotropic scale invariance:
\bea
ds^2 &=& \sqrt{q_1 q_2 q_3 q_4} \left ( - r^2 d \hat{t}^2 + r d \hat{x} \cdot d \hat{x}_2  + \frac{dr^2}{r^2} \right ) \\
&& + \frac{\sqrt{q_3 q_4}} {\sqrt{q_1 q_2}} dy \cdot dy_2 + \frac{\sqrt{q_2 q_4}}{\sqrt{q_1 q_3}} dz \cdot dz_2, \nn
\eea
where $(\hat{t},\hat{x})$ are related to $(t,x)$ by constant rescalings. This observation motivates the idea that anisotropic
scale invariant geometries can be obtained by decoupling limits of intersecting brane systems. However, the actual intersection (\ref{int}) 
used above is not one that actually occurs according to the standard supersymmetric intersection rules. The metric describes
D6-branes intersecting two types of D4-branes over four spatial directions, and D0-branes over time; whilst 
the D4-branes and D0-branes are mutually supersymmetric, the D6-branes are not. It would clearly be interesting to find an  
explicit example of a supersymmetric brane intersection which admits a decoupling limit that gives an anisotropic scale invariant geometry.

\section{Field theories with anisotropic scale invariance} \label{s-field}

Let us consider a $d$-dimensional field theory with anisotropic scale invariance, i.e. it is invariant under scale 
transformations such that $t \rightarrow \lambda^{\alpha} t$ and $x_{i} \rightarrow 
\lambda^{\alpha_i} x_i$. The coefficients $(\alpha, \alpha_i)$ are the dynamical exponents, and
for the most part we will be interested in the case where all $\alpha_i = 1$, and
thus the theory is also invariant under the $(d-1)$-dimensional Euclidean group. A simple model
with this type of scale invariance (with $\alpha = 2$) is the Lifshitz theory:
\be
I = \int dt d^{2} x \left ( (\pa_t \Phi)^2 -  \kappa (\pa^2 \Phi)^2 \right ). 
\ee
This theory is known to have a line of fixed points parameterized by $\kappa$ and arises at a number of finite
temperature critical points in condensed matter systems. 
In this section we will discuss how anisotropic scale invariance constrains correlation functions, Ward identities
and thermodynamic properties. 

The entropy of the field theory should behave extensively in the volume of the $(d-1)$-dimensional space, and therefore
the entropy at finite temperature $T_H$ must scale as
\be
S = c(g^2,\cdots) V_{d-1} T_{H}^{\frac{(d-1)}{\alpha}},
\ee
where $g$ is the dimensionless coupling, the ellipses denote all additional dimensionless parameters (such as the
rank of the gauge group) and $c(g^2,\cdots)$ denotes an arbitrary function of these dimensionless parameters. 

Non-relativistic scale invariance is substantially less restrictive for the two point functions. Two point functions 
of operators of scaling dimension $\Delta$ behave as:
\be
\< {\cal O} (t,x) {\cal O} (t',x') \> = f (\frac{|x-x'|^2}{(t-t')}, g^2, \cdots) \frac{1}{|x - x'|^{2 \Delta}}, 
\ee
where the ellipses again denote additional dimensionless parameters and $f (\frac{|x-x'|^2}{(t-t')}, g^2, \cdots)$ is
an arbitrary function of these dimensionless variables. Time and space translation invariance imply that the correlation
function depends only on $|x-x'|$ and $(t -t')$. In particular, this means that correlators can in principle
admit ultra-local contributions, localized in space or time, and such behavior is of interest in explaining certain
condensed matter observations \cite{Con}. 

Whilst the anisotropic scale invariance leaves considerable freedom in the correlation functions, it implies Ward 
identities for the correlation functions. To derive such relations, one needs to understand how the scale 
invariant action is coupled to background sources. Let us begin by considering the Lifshitz model, and coupling it to
a background diagonal metric via:
\be
I = \int dt d^2x \sqrt{f h} \left ( f^{-1} 
(\pa_t \Phi)^2 -  \kappa (h^{ij} D_{i} \pa_j \Phi)^2 \right ). 
\ee
This Lagrangian is invariant under Weyl transformations such that:
\be
f \rightarrow e^{4 \sigma} f; \qquad
h_{ij} \rightarrow e^{2 \sigma} h_{ij}.
\ee
Let us now define operators such that:
\be
{\cal T} = - \frac{2 }{\sqrt{fh}} \frac{\delta I}{\delta \sqrt{f}}; \qquad
{\cal T}_{ij} = \frac{2}{\sqrt{fh}} \frac{\delta I}{\delta h^{ij}}. 
\ee
These are given by:
\bea
{\cal T} &=&  2 \left ( f^{-1} 
(\pa_t \Phi)^2 +  \kappa (h^{ij} D_{i} \pa_j \Phi)^2 \right ); \\
{\cal T}_{ij} &=& 4 \kappa (h^{kl} D_{k} \pa_l \Phi) D_{i} \pa_j \Phi + h_{ij} (f^{-1} 
(\pa_t \Phi)^2 - \kappa  (h^{ij} D_{i} \pa_j \Phi)^2 ). \nn
\eea 
These clearly satisfy the trace Ward identity:
\be
({\cal T} - {\cal T}^{i}_{i}) = 0, 
\ee
in agreement with the result given earlier in section \ref{s-two}; note however that in general
this identity picks up a conformal anomaly. A more detailed treatment of the
Ward identities implied by the anisotropic scale invariance will be given in \cite{mmt2}, along with
the complete dictionary between these field theory operators and the bulk fields.

\section{Holographic renormalization: scalars in a fixed background} \label{s-scalar}

Both to test and to use the conjectured holographic duality, one needs to set up a precise holographic
dictionary. As usual, one takes the defining holographic relation (at low energy) to be that the 
onshell action with fixed boundary conditions $\phi_{(0)}^{\cal A}$ for 
bulk fields $\phi^{\cal A}$ acts as the generating
functional for connected correlation functions of the dual operators ${\cal O}_{\phi^{\cal A}}$ in the presence
of sources  $\phi_{(0)}^{\cal A}$ \cite{Gubser:1998bc, Witten:1998qj}. 
To render this definition well-defined, one needs to treat systematically the volume divergences of the bulk action,
via holographic renormalization \cite{Henningson:1998gx,Balasubramanian:1999re,deHaro:2000xn,Skenderis:2000in,
BFS1,BFS2,Skenderis:2002wp}.
The matching of the analytic structure of these divergences with the UV 
divergences of the dual theory provides strong structural evidence for the conjectured duality. 
As a warmup case, it is useful to consider scalar field perturbations in the fixed background, and the 
corresponding correlation functions of the dual scalar operators. Systematic holographic renormalization
is necessary to obtain the correct correlation functions, and we will now proceed to develop this. 
Note that earlier discussions of anisotropic holography may be found in \cite{BrittoPacumio:1999sn,Taylor:2000xf}.

\subsection{Regular solutions for scalar fields}

Consider again the simplest non-relativistic background, in
which the $(D+2)$-dimensional metric is
\be
ds^2 = \frac{1}{u^2} \left ( \frac{dt^2}{u^{2\eta}} + du^2 + dx \cdot dx_D
\right ).
\ee
Here we analytically continue to the Euclidean, since in this section we will be interested in
computing correlation functions. Now consider a free massive scalar field in this background, with action
\be
I = \frac{1}{2} \int d^{D+2} x \sqrt{g} (g^{mn} \pa_m \phi \pa_n \phi
+ M^2 \phi^2).
\ee
The field equation can be written as:
\be
u^{D+ \eta} \pa_u ( u^{- (D+ \eta)} \pa_u \phi) + (u^{2\eta} \pa_t^2 + \Box_D) \phi
- \frac{M^2}{u^2} \phi = 0.
\ee
Here $\Box_D$ is the Laplacian in the flat metric on $R^D$. Fourier transforming in both the time
and $R^D$ directions, the equation for ${\phi}(\omega,k)$ becomes:
\be
u^{D+ \eta} \pa_u ( u^{- (D+ \eta)} \pa_u \td{\phi}) - (u^{2\eta} \w^2 + k^2) \td{\phi}
- \frac{M^2}{u^2} {\phi} (\omega,k) = 0.
\ee
In the massless case, the solution which is regular everywhere is:
\be
{\phi} = {\phi}_{(0)}(\omega,k) e^{-\frac{1}{2} \omega x^2} U 
\left (\frac{k^2 + \omega (1 - D - \eta)}{4 \omega}, 
\frac{1}{2} (1 - D - \eta), \omega x^2 \right ), 
\ee
where $U(a,b,x)$ is the Kummer function, normalized such that $U(a,b,0) =1$,
and ${\phi}_{(0)}(\omega,k)$ is arbitrary. In the massive case,
the regular solution is (see also \cite{Kachru:2008yh}):
\bea
{\phi} &=& {\phi}_{(0)}(\omega,k) e^{-\frac{1}{2} \omega x^2} x^{2c} U (a, b, \omega x^2); \label{massive} \\
a &=& \frac{1}{2} - \frac{1}{4} \sqrt{(1 + D + \eta)^2 + 4 M^2} + \frac{k^2}{4 \omega}; \nn \\
b&=& 1 - \frac{1}{2} \sqrt{ (1 + D + \eta)^2 + 4 M^2 L^2}; \qquad
c =  \qu (1 + D + \eta) - \qu \sqrt{ (1 + D + \eta)^2 + 4 M^2}, \nn 
\eea
where we assume that $ 4 M^2 \ge - (1 + D + \eta)^2$; one would anticipate that the latter corresponds
to the Breitenlohner-Freedman bound for this case.

The two point functions for the dual operators were calculated in \cite{Kachru:2008yh} by evaluating
the onshell action, retaining only the finite term, and then functionally differentiating with respect
to the source. Such a procedure of removing the infinities is inconsistent, and in general gives the wrong
answer: in most cases, the one point function of an operator is not simply the normalisable mode of the dual field,
but terms non-linear in the sources can contribute also. 
Here we will derive renormalized one point functions for the operator dual to the free scalar fields.
We will proceed using the standard principles of holographic renormalization \cite{Skenderis:2002wp}: 
we derive the general
asymptotic solution to the field equations, parameterize the volume divergences of the onshell action, 
introduce local counterterms and then obtain the renormalized one point function by functionally differentiating
the renormalized action. 

\subsection{Asymptotic expansion}

Let us begin by considering the general asymptotic expansion of solutions to the scalar field 
equation; these can be written,
as in $AdS_{D+2}$, as
\be
\phi = u^{\Delta_-} (\phi_{(0)}(t,x) + \cdots) + u^{\Delta_+}
(\td{\phi}_{(0)}(t,x) + \cdots),
\ee
where $\Delta_{\pm}$ are the two roots of
\be
\Delta (\Delta - D - 1 - \eta) = M^2,
\ee
and $\Delta_{+} > \Delta_{-}$, $\Delta_{+} = \Delta_{-} + \sqrt{(D+1+ \eta)^2 + 4 M^2}$. 
As usual, there are special cases where the roots are degenerate, namely
when $4 M^2 = - (D + 1 + \eta)^2$, and thus the second solution is logarithmic:
\be
\phi = u^{\Delta} \left (\phi_{(0)}(t,x) + \cdots
+ \ln (u) (\td{\phi}_{(0)}(t,x) + \cdots) \right ). 
\ee
Let us restrict to cases where the roots are not degenerate, as generalizing to the other cases 
is straightforward. In particular, note that in the massless case the normalizable mode scales with
dimension $(D+1 + \eta)$; one would therefore anticipate that the corresponding dual operator is
of this scaling dimension. 

Asymptotically expanding the general solution to the field equation, one finds that the solution takes the form:
\bea
\phi &=& u^{\Delta_-} \phi(t,x,u) + u^{\Delta_{+}} \td{\phi}(t,x,u); \\
\phi &=& u^{\Delta_-} \left (\phi_{(0)}(t,x) + u^2 \phi_{(2)}(t,x) + \cdots \right ) + u^{\Delta_+}
(\td{\phi}_{(0)}(t,x) + \cdots). \nn
\eea
Recall that $\eta \ge 0$; the expansions depend on whether $\eta$ is an integer or a ratio of integers or 
not. In all cases, the first
subleading term in the expansion satisfies the equation: 
\be
(\Delta_- + 2) (\Delta_- - D + 1 - \eta) \phi_{(2)} - M^2 \phi_{(2)} + \Box_D \phi_{(0)} = 0.
\ee
Now consider the case that $0 < \eta < 1$: then the subsequent terms in the expansion are:
\bea
\phi(t,x,u) &=& \phi_{(0)}(t,x) + u^2 \phi_{(2)}(t,x) + u^{2 + 2 \eta} \phi_{(2 + 2 \eta)}(t,x) 
+ u^{2 + 4 \eta} \phi_{(2 + 4 \eta)}(t,x)  + \cdots \nn \\
&& + u^4 \phi_{(4)}(t,x) + \cdots. 
\eea
Here one finds that 
\be
(\Delta_- + 2 + 2 \eta) (\Delta_- + 1 + \eta - D) \phi_{(2 + 2 \eta)} 
+ \pa_t^2 \phi_{(0)} = 0,
\ee
For $1/2 < \eta < 1 $ the next term in the expansion is at order $u^4$ and is determined by:
\be
(\Delta_- + 4) (\Delta_- + 3 - \eta - D) \phi_{(4)} - M^2 \phi_{(4)} +
\Box_D \phi_{(2)} = 0. \nn
\ee
The subsequent terms in the expansion in this case will be at order $(u^{2 + 4 \eta}, u^6, u^{2+ 6 \eta})$ etc.
For $\eta < 1/2$, the ordering of the terms in the expansion is different. 
Note that when $\eta = 1$, the expansion is of the form:
\be
\phi(t,x,u) = \phi_{(0)}(t,x) + u^2 \phi_{(2)}(t,x) + u^4 \phi_{(4)}(t,x) + \cdots,
\ee 
with
\be
(\Delta_- + 4) (\Delta_- + 2 - D) \phi_{(4)} - M^2 \phi_{(4)} + \pa_t^2 \phi_{(0)}
+ \Box_D \phi_{(2)} = 0.
\ee
Clearly there will be special cases for particular choices of $(D,M)$; for example, when $D=2$ and 
$M=0$ one needs a logarithmic term:
\be
\phi(t,x,u) = \phi_{(0)}(t,x) + u^2 \phi_{(2)}(t,x) + u^4 \ln (u) \phi^{l}_{(4)}(t,x) + \cdots,
\ee
with
\be
4  \phi^l_{(4)} (t,x)  + \pa_t^2 \phi_{(0)} + \Box_D \phi_{(2)} = 0.
\ee
More generally, it is straightforward to see when logarithmic terms, related to conformal anomalies, will arise 
in the expansion. In AdS/CFT conformal anomalies for the scalars arise 
whenever $(\Delta_{+} - \Delta_{-})$ is an even integer. 
However, from the asymptotic expansion given here, it is clear there will also be conformal anomalies
when $(\Delta_{+} - \Delta_{-})$ a multiple of $2 \eta$, i.e. when
\be
\Delta_{+} - \Delta_{-} = \sqrt{(D+1+\eta)^2 + 4 M^2} = 2 (k + l \eta), \label{spec-lim}
\ee
where $(k,l)$ are integers. In particular for the massless case there are conformal anomalies when
\be
\eta = \frac{(D+1 - 2k)}{(2l -1)},
\ee
so that for $D$ even $\eta$ is a ratio of odd integers, and for $D$ odd $\eta$ should be a quotient of
an even number by an odd number. Whilst the details of the expansions therefore depend on the case of interest,
in all cases the asymptotic expansions are determined locally in terms of 
the independent non-normalizable and normalizable modes $(\phi_{(0)}, \td\phi_{(0)})$. 

\subsection{Renormalized action and one point functions}

Let us consider the renormalization of the onshell action. The details of the analysis 
will differ depending on the dimension, mass and the specific value of $\eta$, but the defining principles are the same in every case.
Consider first the case of a massless scalar. The onshell action, regulated at $u = \ep$, is given by:
\be
I = \int d^{D+2} x (\sqrt{g} \phi n \cdot \phi)_{u = \ep}
= \int d^{D} x dt (\frac{1}{u^{D+\eta}} \phi \pa_u \phi)_{u = \ep}.
\ee
Evaluating this using the asymptotic expansions given above (for $\eta \le 1$) leads to:
\bea
I = \int d^{D+1} x \left [ \frac{1}{\ep^{D + \eta -1}} 
\left (2 \phi_{(0)} \phi_{(2)} + 2 (1 + \eta) \ep^{2 \eta} \phi_{(0)} \phi_{(2 + 2 \eta)}
+ 2 \ep^2 \phi_{(2)}^2 \right . \right . \\
 \left . \left . + \ep^2 (4 \ln (\ep) + 1) \phi_{(0)} \phi_{(4)}^l + \cdots \right )  + (D+1 + \eta)
\phi_{(0)} \td{ \phi}_{(0)} + \cdots \right ]. \nn
\eea
Given that the expansion coefficients $\phi_{(2n)}$ can be locally expressed in terms of $\phi_{(0)}$, 
the divergent terms can manifestly be expressed locally in terms of the field $\phi$. 
Clearly however the divergences cannot be removed by local
counterterms which are covariant in $(D+1)$-dimensions. Instead, one should require that
the counterterms respect appropriate anisotropic covariance. Note that the asymptotic expansions and regulated
actions even
in these very simple examples are already rather complicated, as the expansion is in both integral powers
of $u$, and in powers of $u^{\eta}$. In systematically developing holography for these cases, it is
therefore important to exploit the more powerful Hamiltonian formalism 
\cite{Papadimitriou:2004ap,Papadimitriou:2004rz}. 

Let us now give the counterterms and renormalized one point functions in two specific but representative
cases: $D =2$ with $1/2 < \eta < 1$ (and thus not a ratio of odd integers) 
and $\eta = 1$ respectively. In the first case appropriate counterterms are
\be
I_{ct} =  - \int d^{3}x \sqrt{h} \left ( \frac{1}{(\eta + 1)} \phi \Box \phi + \frac{1}{(1 - \eta)} 
\phi D^{t} \pa_t \phi \right ), 
\ee
and thus the renormalized one point function for the operator ${\cal O}_{\eta} $ dual to $\phi$ is
\be
\< {\cal O}_{\eta} \> \equiv \frac{\delta I_{ren}}{\delta \phi_{(0)}} = (3 + \eta) \td{\phi}_{(0)}.  
\ee
In the second case one needs in addition logarithmic counterterms so that
\be
I_{ct} = - \int d^{3}x \sqrt{h} \left ( \frac{1}{2} \phi \Box \phi + \ln (\ep)
( \phi D^{t} \pa_t \phi + \qu \Box^2 \phi) \right ),
\ee
and the renormalized one point function for the operator ${\cal O}_{1} $ dual to $\phi$ is
\be
\< {\cal O}_{1} \> \equiv  \frac{\delta I_{ren}}{\delta \phi_{(0)}}
= 4 \td{\phi}_{(0)} - \frac{3}{8} \Box^2 \phi_{(0)} - \frac{1}{2} D^{t} \pa_t \phi_{(0)}. 
\ee
Note that the expressions given in \cite{Kachru:2008yh} correspond to retaining only the term linear
in the normalizable mode, and not the terms involving the source (non-normalizable mode). These 
additional terms contribute contribute contact terms to the higher point functions; if we are interested
in whether the correlation functions admit ultra-local behavior at finite temperature, in general, we should keep track
of all such local terms. 

One can next compute the two point functions, using the expressions for the exact, regular solutions; 
in both cases the result agrees with the expectations given the anisotropic scale invariance. In the
first case one finds:
\be
\< {\cal O}_{\eta} (\omega, k) {\cal O}_{\eta} (-\omega, -k)\> =   
- (3 + \eta) \omega^{(3 + \eta)/2} \frac{ \Gamma[- \frac{1}{2} (3  +\eta) ]
\Gamma[ \frac{k^2}{4 \omega}  
+ \frac{(5 + \eta)}{4}]}{\Gamma[\frac{1}{2} (3  + \eta)] \Gamma[ \frac{k^2}{4 \omega}  - \frac{1}{4} (1 + \eta)]}. 
\ee
In the second case the expression is rather more complicated. The asymptotic expansion of
the regular solution is:
\bea
{\phi}(\omega,k) &=& {\phi}_{(0)}(\omega,k) e^{-\frac{1}{2} \omega u^2} U 
\left (\frac{k^2 - 2 \omega}{4 \omega}, 
-1 , \omega u^2 \right ); \nn \\
&=& {\phi}_{(0)}(\omega,k) \left (1 - \frac{k^2}{4} u^2 +  (3k^4 - 20 \omega^2 + 2\gamma(4\omega^2 − k^4 ))  
\frac{u^4}{64} \right . \\
&& \qquad \left . + \frac{u^4}{64} (8 k^2 \omega + (4 \omega^2 - k^4) \ln (\omega^2 u^4) + 2 (4 \omega^2 - k^4)
\psi(\frac{3}{2} + \frac{k^2}{4 \omega}) ) + \cdots \right ). \nn 
\eea
where $\psi(x) = \Gamma'(x)/\Gamma(x)$ is the digamma function and $\gamma$ is the Euler-Mascheroni
constant. From this expression one can determine that 
\bea
\td{\phi}_{(0)} &=& \frac{1}{64} {\phi}_{(0)}
\left (3k^4 - 20 \omega^2 + 2\gamma(4\omega^2 − k^4 )  + 8 k^2 \omega \right . \\
&& \qquad \left . + (4 \omega^2 - k^4) \ln (\omega^2) + 2 (4 \omega^2 - k^4)
\psi(\frac{3}{2} + \frac{k^2}{4 \omega}) \right ), \nn
\eea
and thus that the two point function is:
\bea
\< {\cal O}_{1} (\omega, k) {\cal O}_{1} (-\omega, -k)\> &=&
- \frac{1}{16} \left (3k^4 - 20 \omega^2 + 2\gamma (4\omega^2 − k^4 )  + 8 k^2 \omega
+ (4 \omega^2 - k^4) \ln (\omega^2) \right . \nn \\
&& \qquad \left . + 2 (4 \omega^2 - k^4)
\psi(\frac{3}{2} + \frac{k^2}{4 \omega}) \right ) + \frac{3}{8} k^4 - \frac{\omega^2}{2}.
\eea
This expression differs from that given in \cite{Kachru:2008yh}, only by the contact terms. 

\bigskip

Next let us consider the case of a massive scalar; for simplicity let us again restrict
to $D=2$ and $1/2 < \eta < 1$ with the mass not taking the special values given in (\ref{spec-lim}). 
Repeating the same steps as before, the regulated onshell action is:
\bea
I &=& \int d^{3} x \left [ \frac{1}{\ep^{3 + \eta - 2 \Delta_-}} 
\left (\Delta_{-} \phi_{(0)}^2 + 2 (1 + \Delta_{-}) \ep^2 \phi_{(0)} \phi_{(2)} + 2 (1 + \eta + \Delta_{-}) 
\ep^{2 + 2 \eta} \phi_{(0)} \phi_{(2 + 2 \eta)} + \cdots \right ) \right . \nn \\
&&  \left . + (D+1 + \eta)
\phi_{(0)} \td{ \phi}_{(0)} + \cdots \right ]. 
\eea
Appropriate counterterms are
\be
I_{ct} =  - \int d^{3}x \sqrt{h} \left ( \Delta_{-} \phi^2  + a_1 \phi \Box \phi + a_2 \phi D^{t} \pa_t \phi \right ), 
\ee
where
\be
a_1 = - \frac{2}{(\Delta_- + 2)(\Delta_{-} -1 - \eta) - M^2}; \qquad
a_{2} = - \frac{2 (1+ \eta)}{(\Delta_{-} + 2 (1 + \eta))(\Delta_{-} + \eta -1)}. 
\ee
The renormalized one point function for the dual operator is then
\be
\< {\cal O}_{\Delta} \> \equiv \frac{\delta I_{ren}}{\delta \phi_{(0)}} = (D + 1 + \eta - 2 \Delta_{-}) \td{\phi}_{(0)};  
\ee
there are no local contributions for these generic values of $M$ and the expression is written in a form that
is appropriate for all dimensions $D$ (see \cite{mmt2}). 
Note that this expression coincides with that given in \cite{Skenderis:2002wp} in the limit that $\eta = 0$. 
Extracting the appropriate term from the expansion of the exact regular solution (\ref{massive}) gives a two
point function which is 
\bea
\< {\cal O}_{\Delta} (\omega, k) {\cal O}_{\Delta} (-\omega, -k)\> = - \sqrt{ (D +1 + \eta)^2 + 4M^2} 
\omega^{\half \sqrt{(D +1 + \eta)^2 + 4M^2}} \\ 
 \frac{\Gamma(-\frac{1}{2} \sqrt{(D +1 + \eta)^2 + 4M^2})}
{ \Gamma(\frac{1}{2} \sqrt{(D +1 + \eta)^2 + 4M^2})} \frac{\Gamma (\frac{k^2}{4 \omega} + \frac{1}{2} + \qu \sqrt{ (D+1+ \eta)^2 + 4 M^2})}
{\Gamma (\frac{k^2}{4 \omega} + \frac{1}{2} - \qu \sqrt{ (D+1+ \eta)^2 + 4 M^2})}. \nn 
\eea
Restricting to the case of $\eta = 2$, $D=2$, which was given \cite{Kachru:2008yh}, one sees that the overall
normalization is different. This was to be expected: in the AdS case it is well-known that the prescription of 
\cite{Witten:1998qj} gives the wrong normalization, except for operators dual to massless fields. In all other cases, the counterterms
needed to render the action finite also give contributions to the renormalized correlation functions, see \cite{Skenderis:2002wp}. 
Without carrying out
systematic holographic renormalization, one will obtain inconsistent normalizations, and the expected Ward identities will not
be satisfied. Moreover, as was demonstrated in 
\cite{BFS1,BFS2,Skenderis:2002wp}, once one moves away from the scale invariant point to consider RG flows, correlators obtained
by inconsistent subtraction schemes give not just inconsistent normalizations but qualitatively incorrect physics. 

The renormalized one point functions derived here for scalars in a fixed background suffice to compute finite temperature correlation
functions of the corresponding operators. The fact that the divergences in this
case are local, and can be removed by local counterterms, indeed provides supporting structural evidence for the duality. A much stronger test,
however, will be provided by carrying out the same procedure in the full non-linear system involving the metric and matter fields and will
be developed in \cite{mmt2}.   

\section*{Acknowledgments}

The author is supported by NWO, via the Vidi grant ``Holography,
duality and time dependence in string theory''. This work was also
supported in part by the EU contract MRTN-CT-2004-512194.

\end{document}